# SYSTEMATICS OF HEAVY QUARK PRODUCTION AT HERA[*]

**Stanley J. Brodsky and Wai-Keung Tang**

*Stanford Linear Accelerator Center*

*Stanford University, Stanford, California 94309*

and

**Paul Hoyer**

*NORDITA, Copenhagen, Denmark*

ABSTRACT

We discuss heavy quark and quarkonium production in various kinematic regions at the HERA $ep$ collider. In contrast to fixed target experiments, collider kinematics allows the possibility of detailed measurements of particle production in the proton fragmentation region. One thus can study parton correlations in the proton Fock states materialized by the virtual photon probe. We discuss various configurations of inelastic electron-proton scattering, including peripheral, diffractive, and deep inelastic processes. In particular, we show that intrinsic heavy quark Fock states can be identified by the observation of quarkonium production at large $x_F$ and a low mean transverse momentum which is insensitive to the virtuality $Q^2$ of the photon.

Submitted to Physical Review **D**.

[*]Work supported in part by Department of Energy contract DE–AC03- -76SF00515 and DE–AC02–76ER03069.

# 1 Introduction

The structure of hadrons at short distances is a central focus of studies in QCD. Most existing information on single quark and gluon distributions in nucleons comes from measurements of the structure functions in deep inelastic lepton scattering. The advent of the HERA electron-proton collider opens up possibilities for detailed studies of the target (proton) fragmentation. In a typical deep inelastic scattering event, a virtual photon with transverse polarization in effect removes the struck quark from its proton Fock state. The ensuing proton fragmentation, *i.e.*, the spectator jet, then reveals *correlations* between the struck quark and other partons in the same light-cone Fock state of the proton. Such correlations cannot be inferred from fixed target measurements which only provide information on the scattered lepton and the hadrons produced in the current jet.

A study of correlations between the scattered lepton and particles produced in the proton fragmentation region was recently suggested by Trentadue and Veneziano [1]. They emphasized the study of the evolution of "fracture functions"; *i.e.*, the change in the momentum distributions of the target fragments as a function of the momentum transfer of the lepton probe. Here we shall focus on the physics of the proton itself as illuminated by target fragmentation. In general, the proton's fragments are produced within a few units of rapidity of the incident proton and thus appear at HERA as particles with a large fraction of the proton's momentum. In particular, we shall show that the correlations between the struck quark and the hadrons containing heavy quarks in the proton fragmentation region provide a new type of microscope for resolving the short-distance structure of the proton's Fock state wavefunction. A heavy quark can be experimentally tagged by the heavy meson which it forms, since the meson and quark velocities are nearly the same. A heavy quark-antiquark pair with similar rapidities in the proton wavefunction can be tagged by observing heavy quarkonium in the proton fragmentation region.

Electron-proton scattering allows the hardness of the virtual photon probe to be precisely monitored. Next-to-leading order calculations at leading twist have recently been carried out by Harris and Smith [2], who computed heavy quark production in deep inelastic $ep$ scattering, and by Krämer *et al.*[3], who considered inelastic $J/\psi$ photoproduction. We shall discuss more generally the physics of heavy quark production in both deep inelastic scattering (including higher twist effects) and in peripheral scattering. In the latter case, either the proton or the electron emerges in the final state with only a small change of momentum.

We shall argue that experiments at HERA can detect unusual Fock states in the proton, closely related to "intrinsic charm" [4], since these wavefunction configurations will give rise to heavy quark production in an unusual kinematic region. The primary source of heavy quarks in the proton wavefunction at low light-cone momentum fraction $x = k^+/p^+ = (k^0 + k^z)/(p^0 + p^z)$ is gluon splitting $g \to Q\overline{Q}$. However, heavy quarks are also created from subprocesses involving multiple connections to the other constituents such as $gg \to Q\overline{Q}$. Such intrinsic heavy quark components are correlated with the bound state structure of the proton, not just the structure of the gluon. Since a bound-state wavefunction favors configurations with minimal off-shell energy and minimal invariant mass of the constituents, the intrinsic heavy quark momentum fraction $x_Q$ tends to peak at relatively large momentum fraction. The EMC measurements of the charm structure function extracted from $\mu p \to \mu \mu X$ data show an excess over predictions based on gluon splitting and suggest that the probability $P_{c\bar{c}}$ of intrinsic charm in the proton is of order 0.3% [5]. Theoretical estimates [6] suggest that $P_{c\bar{c}}$ is of order 1%. Since the intrinsic probability scales as $1/m_Q^2$, the probability for intrinsic bottom is expected to be one order of magnitude smaller. In this analysis we shall show how measurements of proton fragmentation at HERA can discriminate between the gluon-splitting and intrinsic sources of heavy quarks in QCD.

## 2 Heavy Quark Production in $ep$ collisions

Consider first deep inelastic scattering (DIS) in which the photon virtuality $Q^2 \gg M^2$, where $M/2$ is the mass of the heavy quark. There are at least two jets in the final state – the current (virtual photon) jet and the target (proton) jet. We shall assume that only a single heavy quark-antiquark pair is produced in the event. We then distinguish three classes of heavy quark events, according to whether the heavy quark and antiquark appear in different jets, or if both are in the current jet, or if both are in the target jet. In the following we shall for definiteness assume that the heavy quark is a charm quark.

### 2.1 DIS with the charm quark and antiquark in separate jets.

This is the typical configuration for charm production in DIS in which the virtual photon resolves the $c\bar{c}$ fluctuation in a gluon when $Q^2 \gg M^2$ (Fig. 1(a)).[†] The incident lepton gives a large momentum transfer to the struck heavy quark whose light-cone momentum fraction in the proton is fixed by the scattered lepton kinematics to be $x_{Bj} = Q^2/2p \cdot q$. The contribution to the charm structure function which arises from gluon splitting, or QCD evolution, scales as $\alpha_s \log Q^2/m_c^2$. (For definiteness, we shall assume that the struck heavy quark is the $c$ quark.) The struck charm quark fragments into a charmed hadron (typically a $D$ meson) which appears in the current jet carrying a large fraction $z$ of the photon energy. According to the QCD factorization theorem [7], the charmed jet fragmentation function $D_{c \to D}(z)$ is universal and thus should agree with the fragmentation function measured at comparable scales in $e^+e^- \to D + X$. The charmed antiquark spectator makes an anticharm (say, $\overline{D}$) hadron in the target (proton) jet. At leading twist, the $\bar{c} \to \overline{D}$ hadronization should be described by the same universal fragmentation function $D_{c \to D}(z)$.

---

[†]Equivalently, we may say that the hard subprocess is (virtual) photon - gluon fusion, $\gamma^* g \to c\bar{c}$.

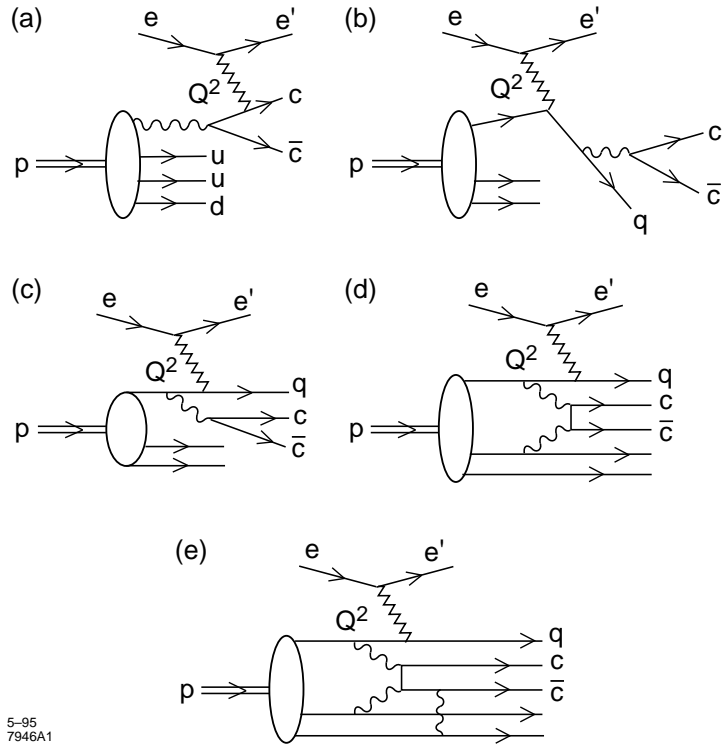

Figure 1: DIS heavy quarks production with $Q^2 \gg M^2$. (a) The charm quark and antiquark in separate jets. (b) Both the charm quark and antiquark in the photon jet. (c) Both the charm quark and antiquark in the proton fragmentation region. (d) Higher twist contribution where two gluons are coupled to the charm quark and antiquark. (e) Three gluons coupled to the charm quark and antiquark with the quantum numbers of the $J/\psi$.

The data on the hadroproduction of $D$ mesons shows that there actually are considerable differences between the charm quark hadronization in the fragmentation region of a hadron, as compared to that measured in $e^+e^-$ annihilation [8]. The difference is presumably due to the interactions between the produced charm quark and the co-moving spectator partons from the incident hadron. This is indicated by the existence of very strong factorization-breaking quantum number correlations between the forward-produced $D$ mesons and the projectile hadron [9]. In charm photoproduction, on the other hand, where the directly-interacting photon leaves no co-movers, charm quark hadronization is consistent [10] with that measured in $e^+e^-$ annihilation.

Co-mover interactions allow partons of similar rapidities to coalesce into hadrons. The dominant production of charmed quarks is at fairly low $p_T = \mathcal{O}(m_c)$ relative to the beam direction. It is thus not surprising to find strong leading particle correlations of the charge of the produced $D$ mesons with the quantum numbers of the beam at large values of the fractional momentum $x_F$ of the $D$ mesons [11] where the charm quark and the spectator fragments are moving at similar rapidities. The co-mover interactions between the produced charmed quark and the spectator partons in the projectile fragmentation region is a higher-twist QCD effect which breaks leading-twist PQCD factorization [12]. The induced correlations from the co-mover interactions are suppressed at large transverse momentum by powers of the hard scale.

Because of the breakdown of factorization, heavy quark hadronization in the proton fragmentation region in DIS need not be the same as that in heavy quark hadroproduction. However, it is likely that to a first approximation the effective fragmentation function will be similar to that observed [8] in $pN \to D + X$, i.e., $D_{\bar{c} \to \overline{D}}(z) \simeq \delta(1-z)$.

If the momentum of a $\overline{D}$ meson in the proton jet can be measured, one has thus a good estimate of the original momentum of its charm antiquark constituent. Since the momentum of its charm quark partner is given by $x_{Bj}$, the fractional momentum $x_{c\bar{c}}$ initially carried by the $c\bar{c}$ pair in the proton can be estimated. To leading order in $1/m_c^2$, the charm quark pairs are produced through the splitting of single gluons. Thus their momentum distribution should agree with that given by the gluon distribution function $F_{g/p}(x_g)$, with $x_{c\bar{c}} = x_g$. The accuracy of this prediction may be checked by verifying that the fraction $z_c$ of the charm quark pair momentum carried by one charm quark reflects the gluon splitting function $P_{g \to c\bar{c}}(z_c) = [z_c^2 + (1-z_c)^2]/2$. The ZEUS Collaboration [13] recently observed the production of $D^*(2010)$ in DIS events at HERA. The published data, however, does not distinguish whether the charmed mesons belong to the photon or proton jets.

## 2.2 DIS with both the charm quark and antiquark in the photon jet.

The dominant production process at large $Q^2$ for this configuration is illustrated in Fig. 1(b). The photon scatters off a light quark, which subsequently fragments into a $c\bar{c}$ pair. The probability for the struck quark to fragment into charmed quarks scales as $\alpha_s \log Q^2/m_c^2$. The dominance of fragmentation diagrams is analogous to charmonium production at large $p_\perp$ in hadron collisions [14]. Contributions of the type shown in Fig. 1(a), where the charm quark pair is present in the initial proton Fock state, require a momentum transfer between the charm quarks of order $Q$ to turn both quarks in the photon direction. Such contributions are suppressed by powers of $M^2/Q^2$.

The $c\bar{c}$ pair can most easily be detected experimentally as a charmonium state. For S (P) wave charmonia, two (one) gluons need to be emitted from the charm quarks in order to give the right quantum numbers to the pair. The strong disagreement recently found between leading-twist QCD predictions for color-singlet $c\bar{c}$ bound states [15] and the CDF data [16] for charmonium and bottomonium production in hadron collisions makes a study of quarkonium production in DIS very important. As in the case of $e^+e^-$ annihilation, the initial momentum of the fragmenting quark is known in DIS. However, in $e^+e^-$ a dominant contribution is expected to come from the case where the primary fragmenting quark is itself a charm quark [15]. In $ep$ collisions this is suppressed due to the relatively small magnitude of the charm quark structure function of the proton.

## 2.3 DIS with both the charm quark and antiquark in the proton fragmentation region.

This is perhaps the most interesting of the DIS charm event configurations. In order to avoid giving a large transverse momentum of order $Q$ to the charm quarks, the lepton must scatter off a *light* quark in the proton as illustrated in Figs. 1(c) and 1(d). Normally, the charm pair will not materialize in the final state even if that struck light quark is part of a proton Fock state $|uudc\bar{c}\rangle$ which contains charm. There is typically no coherence between the light quarks in the wave function and that of the $c\bar{c}$ pair, since the latter is a high energy localized excitation, and the phase of its amplitude changes rapidly with time. There is an exception to this rule if the virtual photon hits the light quark very soon after it has emitted a virtual $c\bar{c}$ pair (Fig. 1(c)). The $q \to qc\bar{c}$ transition involves an energy difference

$$\Delta E = \frac{1}{2E_p} \left[ M_p^2 - \frac{M_{c\bar{c}}^2 + p_\perp^2}{x_{c\bar{c}}} - \frac{M_{\text{spectator}}^2 + p_\perp^2}{1 - x_{c\bar{c}}} \right] \tag{1}$$

where $M_{c\bar{c}}$, $p_\perp$ and $x_{c\bar{c}}$ are the mass, transverse momentum and momentum fraction of the $c\bar{c}$ pair, and $M_{\text{spectator}}$ is the invariant mass of the remaining light quarks. If the photon interacts with one of the light quarks in the $|qc\bar{c}...\rangle$ state within the lifetime of the $c\bar{c}$ fluctuation. Then the typical energy $(M^2_{\text{spectator}} + p^2_\perp)/(1 - x_{c\bar{c}})2E_p$ of the light quark spectator system in Eq. (1) is similar to the energy of the heavy pair. At intermediate values of $x_{c\bar{c}}$, this means that $Q^2 > p^2_\perp \sim M^2_{\text{spectator}} \gtrsim M^2_{c\bar{c}}$. The diagram in Fig. 1(c) is then part of the perturbatively-generated leading-twist logarithmic evolution of the struck parton in DIS where $p^2_\perp$ ranges up to the resolution scale $Q^2$. This type of charm pair production is thus included in the evolution of the proton fracture function [1].

The light quark system will also have an energy compatible with that of the $c\bar{c}$ pair if $(1 - x_{c\bar{c}})M^2_{c\bar{c}} \sim p^2_\perp$ with $p_\perp$ limited and independent of $Q^2$. This condition corresponds to the production of charmonium states carrying a large fraction $x_F = x_{c\bar{c}}$ of the proton momentum with limited $p_\perp$. This kinematic region is evidently suppressed by phase space, since the ranges of both $p_\perp$ and $x_{c\bar{c}}$ are constrained. On the other hand, such configurations minimize the off-shell energy of the proton wavefunction and thus can be dynamically enhanced. In fact, at large $x_F$, higher twist contributions are important [17] and can significantly boost the cross section. An example of such a contribution, which involves several partons in the proton, is shown in Fig. 1(d). The probability of finding the heavy quarks in the proton Fock states from diagrams such as Fig. 1(d) scales nominally as $\mu^2/M^2$ where $\mu$ is a hadronic scale derived from extra loop integration. The fact that three gluons can couple to the charm quark pair within the proton means that the $c\bar{c}$ pair can be produced as a color singlet with the right quantum numbers to form quarkonium without the emission of further gluons (Fig. 1(e)). The large magnitude of the charmonium hadroproduction cross section, the behavior of the $J/\psi$ polarization, and the observed nuclear target $A$-dependence at large $x_F$ can all be understood [19], if such higher twist mechanisms dominate heavy quark production for $x_F \sim 1$.

In $ep$ collisions, the control over the virtual photon probe allows a detailed study of the importance of intrinsic heavy quark contributions such as those shown in Figs. 1(d) and 1(e). In such processes the virtual photon simply materializes the pre-existing large $x_F$ heavy quarks in the proton's Fock state, typically by interacting with a light quark of light-cone momentum fraction $x = x_{Bj} \propto 1 - x_F$. At large $Q^2 \gg M^2$, the cross section $d\sigma/dQ^2 dM^2$ for charm production in the large $x_F$ proton fragmentation region from the higher twist intrinsic diagrams such as Fig. 1(d) scales as $\mu^2/Q^4 M^4$. The heavy quark system appearing at large $x_F$ should have uncharacteristically small $p_\perp \ll M$, and this transverse momentum should not grow with $Q^2$. In hadroproduction there is evidence that the average transverse momentum of $J/\psi$ particles does indeed decrease with $x_F$ [20]. Since the heavy quarks in the proton Fock state tend to carry most of the proton's momentum, the falloff of the cross section at large $x_F$ is less damped than that of the leading-twist contributions which fall at least as fast as the gluon distribution in the proton. In the case of open charm production, the kinematics of the heavy quark system will reflect the characteristics of the proton Fock state. Thus the total transverse momentum of the charmed hadron pair will be small and the invariant mass of the pair will tend to be at the lowest possible values, independent of the virtuality of the photon probe. In contrast, the total transverse momentum and invariant mass of heavy quark systems produced via leading-twist subprocesses will grow with $Q^2$.

Next we turn to soft scattering, $Q^2 \ll M^2$. We distinguish two cases, based on the size of the fraction $y = q \cdot p / k \cdot p$ that the photon (of momentum $q$) carries of the electron momentum $k$ (here $p$ is the proton momentum).

## 2.4 Low $Q^2$, moderate $y$.

This process is effectively photoproduction (Fig. 2(a)). The dominant leading-twist mechanism at high energy is of the Bethe-Heitler type. The almost real photon fluctuates into a $c\bar{c}$ pair which is put on mass shell by a gluon from the target. Charm photoproduction has been studied both in fixed target experiments [10] and at HERA [21]. There is good agreement with leading-twist QCD calculations based on photon-gluon fusion.

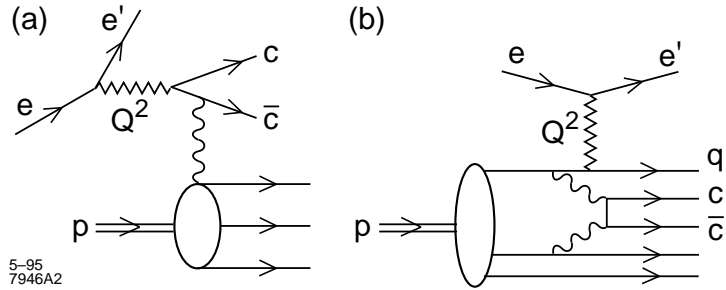

Figure 2: Soft scattering heavy quark production with $Q^2 \ll M^2$. (a) Photoproduction where the charm quark and antiquark in the photon jet. (b) Peripheral scattering where both the electron and proton suffer little momenta loss.

In the case of charmonium production, an important part of the cross section at small $Q^2$ is observed to be elastic or diffractive [22, 23]. Such hard elastic and diffractive production of heavy quarks is of great interest, and should be calculable in perturbative QCD using two gluon exchange [24, 25]. The small transverse size of the heavy quark state implies that both gluons coupling to the charmonium state carry significant transverse momentum even though the overall momentum transfer is small. Recent HERA data [26] for the $\gamma p \to J/\psi p$ cross section shows a rapid rise with energy consistent with the increase of the square of the gluon distribution at small $x \simeq M_{J/\psi}^2/s$ measured in DIS. The data [23] on the nuclear target $A$ dependence of these processes also gives support to the perturbative two-gluon reaction mechanism. The importance of elastic and diffractive scattering of charmonium, as compared to open charm production, is a good example of how the quantum numbers of specific (and rare) final states affects the reaction mechanism, enhancing the contribution from higher-twist processes involving several partons from one of the colliding hadrons.

## 2.5 Low $Q^2$, low $y \sim M^2/s$.

In this case, the electron scatters peripherally off the proton, suffering little momentum loss. In such collisions one normally does not expect even at high energy to excite heavy degrees of freedom such as charm. The reason for this is that a soft momentum exchange cannot resolve a transversely compact heavy quark state.

Soft scattering can, however, occur when the heavy quarks in a Fock state are coherent with light quarks which have a broad transverse size distribution [17] (Fig. 2(b)). The heavy quarks can then be put on their mass shell when the electron scatters on light quarks, as in the DIS case (c) above. If the charm pair takes most of the proton's longitudinal momentum ($x_F \to 1$ for the charmonium state), then the light quarks will have small longitudinal momenta and small balancing transverse momenta $p_\perp^2 \sim M^2(1 - x_F)$. This state can be resolved by a photon with virtuality $Q^2 \simeq p_\perp^2 \sim M^2(1 - x_F)$. Thus in the large $x_F$ regime, production of charmonium and open charm should persist down to $Q^2 \simeq M^2(1 - x_F)$. The net effect will be to make the $x_F$-distribution of charmonium flatter at low $Q^2$. In this regime with $Q^2 \ll M^2$, the cross section $d\sigma/dQ^2 dM^2$ scales as $\mu^2/Q^2 M^6$ since there is an extra $Q^2/M^2$ suppression due to resolution of the charm quark pair [17].

## 3 The Physics of Intrinsic Heavy Quark Fock States

The examples given in the above discussion show that $ep$ collisions at HERA can provide many new insights into the dynamics of heavy quark production. Of particular interest is the possibility to resolve and detect intrinsic heavy quark Fock states [4] in which charm and beauty states pre-exist in the proton projectile. Examples of such higher twist contributions are shown in Figs. 1(d) and 2(b). The most distinctive characteristic of the intrinsic contri-butions is that the average transverse momentum $p_\perp$ of the produced heavy quark system does not increase with the virtuality $Q^2$ of the photon.

There are many contrasts between the leading twist and higher twist mechanisms for heavy quark production in $ep$ scattering. In a leading-twist contribution the photon probes heavy quarks which are fluctuations of a pointlike gluon in the proton. The relative probability of various $c\bar{c}$ fluctuations in a single gluon is only logarithmically dependent on the relative transverse momentum of the charm quarks – this is the definition of a "pointlike" gluon. The higher the $Q^2$ of the photon, the more short-lived fluctuations of the gluon can be resolved. Hence the $p_\perp^2$ of the produced quarks extends up to a scale proportional to $Q^2$.

The heavy quarks in an intrinsic heavy quark Fock state such as in Figs. 1(d), 1(e), and 2(b) have a transverse momentum distribution which is controlled by the proton's light-cone wavefunction. The nominal probability for finding a heavy quark pair generated by a two-gluon diagram falls off as $P_{Q\bar{Q}} \sim \mu^2/M_{Q\bar{Q}}^2$ where $\mu$ is a characteristic hadronic scale of the proton wavefunction and $M_{Q\bar{Q}}$ is the invariant mass of the heavy quark pair. This scaling also characterizes the relative magnitude of the intrinsic charm and beauty structure functions at $Q^2 \gg M_Q^2$. The transverse distance between the heavy quarks in an intrinsic heavy quark Fock state is inversely proportional to the heavy quark mass, whereas the characteristic transverse size of the associated light quarks is of order $1/M\sqrt{1-x_{c\bar{c}}}$, where $x_{c\bar{c}}$ is the fractional momentum of the heavy pair. Thus the intrinsic heavy quark state has a definite probability and size. A photon scattering off one of the light quarks in the intrinsic state can materialize the heavy quarks, but it does not affect their transverse momenta since the transverse distance between the light and heavy quarks is relatively large. Thus the observation of heavy quark systems at large $x_F$ in the proton fragmentation region at HERA should provide a clear signal of intrinsic charm and bottom Fock components in the proton if the transverse momentum of the heavy quark system is insensitive to the virtuality of the photon.

We have emphasized that the longitudinal momentum distributions in the intrinsic heavy quark Fock states are characterized by configurations in which all constituents having similar velocities since the energy of the state is minimized. When such states are materialized in $ep$ collisions the resulting hadrons produced in the proton fragmentation region will tend to have similar rapidities. Luke, Manohar, and Savage [27] have used the operator product expansion and the trace anomaly to show that a strong attractive scalar interaction will arise between quarkonium states and ordinary hadrons with small relative velocity. This is the analog of the attractive Van der Waals interaction which binds atoms into molecules in Abelian QED. In the case of $ep \to e'J/\psi HX$, the intrinsic charm Fock state emphasizes the production of the $J/\psi$ and other hadrons with similar velocities in the proton fragmentation region. One can thus study the $J/\psi H$ system as a function of its invariant mass and search for resonances or even bound states of the $c\bar{c}$ with light hadrons [28]. The study of quarkonium production and correlations in the fragmentation region of the proton in inelastic lepton scattering thus not only probes the short-distance structure of hadron wavefunctions but it also provides a window to new types of QCD interactions involving hadrons at low relative rapidity, including the possible resonance formation of the $J/\psi, \psi'$ and other quarkonium states with ordinary hadrons.

# 4 Acknowledgments

We would like to thank Gunnar Ingelman, Al Mueller, Johan Rothsman, Mikko Vänttinen, and Ramona Vogt for helpful conversations.